\documentclass[preprintnumbers, prd, showpacs, floatfix,onecolumn,
preprintnumbers, letterpaper,
superscriptaddress,nofootinbib]{revtex4}
\usepackage{amsfonts}
\usepackage{amsmath}
\usepackage{amssymb,epsf}
\usepackage{latexsym}
\usepackage{graphicx,epsfig}
\usepackage{amssymb}
\usepackage{subfigure}

\begin{document}

\title{Thermodynamics of charged rotating dilaton black branes \\
with power-law Maxwell field}
\author{M. Kord Zangeneh}
\email{mkzangeneh@shirazu.ac.ir}
\affiliation{Physics Department and Biruni Observatory, College of Sciences, Shiraz
University, Shiraz 71454, Iran}
\author{A. Sheykhi}
\email{asheykhi@shirazu.ac.ir}
\author{M. H. Dehghani}
\email{mhd@shirazu.ac.ir}
\affiliation{Physics Department and Biruni Observatory, College of Sciences, Shiraz
University, Shiraz 71454, Iran}
\affiliation{Research Institute for Astronomy and Astrophysics of Maragha (RIAAM), P.O.
Box 55134-441, Maragha, Iran}

\begin{abstract}
In this paper, we construct a new class of charged rotating dilaton black
brane solutions, with complete set of rotation parameters, which is coupled
to a nonlinear Maxwell field. The Lagrangian of the matter field has the
form of the power-law Maxwell field. We study the causal structure of the
spacetime and its physical properties in ample details. We also compute
thermodynamic and conserved quantities of the spacetime such as the
temperature, entropy, mass, charge, and angular momentum. We find a
Smarr-formula for the mass and verify the validity of the first law of
thermodynamics on the black brane horizon. Finally, we investigate the
thermal stability of solutions in both canonical and grand-canonical
ensembles and disclose the effects of dilaton field and nonlinearity of Maxwell field
on the thermal stability of the solutions. We find that for $\alpha \leq 1$, charged
rotating black brane solutions are thermally stable independent of the
values of the other parameters. For $\alpha>1$, the solutions can encounter
an unstable phase depending on the metric parameters.
\end{abstract}

\maketitle

\section{INTRODUCTION\label{Intr}}

Historically, the presence of a scalar field in the context of general
relativity dates back to Kaluza-Klein theory \cite{KK}. Kaluza-Klein theory
was presented in order to give a unified theory describing gravity and
electromagnetic. It may be considered as a pioneering theory to the string
theory. Afterwards, in Brans-Dicke (BD) theory, the notion was seriously
taken into account \cite{BD}. The root of this theory can be traced back to
Mach's principle encoded by a varying gravitational constant in BD theory.
The variety of gravitational constant is indicated in this theory by
considering a scalar field non-minimally coupled to gravity. BD theory is
known as a non-quantum scalar-tensor theory \cite{B}. From quantum
viewpoint, the scalar field called dilaton field is emerged in the low
energy limit of string theory \cite{Wit1}.

On the other side, there are strong evidences on the observational sides
that our Universe is currently experiencing a phase of accelerating
expansion \cite{Expan}. This acceleration phase cannot be explained through
standard model of cosmology which is based on the Einstein theory of general
relativity. One way for explanation of such an acceleration is to add a new
unknown component of energy, usually called \textquotedblleft dark
energy\textquotedblright . Another way for challenging with this problem is
to modify the Einstein theory of gravity. In this regards, some physicists
have convinced that Einstein theory of gravitation cannot give a complete
description of what really occurs in our Universe and one needs an
alternative theory. As we mentioned above, string theory, in its low energy
limit, suggests a scalar field which is nonminimally coupled to gravity and
other fields called dilaton field \cite{Wit1}. Consequently, dilaton gravity
as one of the alternatives of Einstein theory have encountered high interest
in recent years. String theory also suggests dimensions of spacetime to be
higher than four dimensions. It seemed for a while that dimensions higher
than four should be of Planck scale, but recent theories expresses that our
three-dimensional brane can be embedded in a relatively large higher
dimensional bulk which is still unobservable \cite{RS,DGP}. In such a
scenario, all gravitational objects including black objects are higher
dimensional. The action of dilaton gravity usually includes one or more
Liouville-type potentials which can be justified as trace of supersymmetry
breaking of spacetime in ten dimensions.

Many authors have explored exact charged black solutions in the context of
Einstein-dilaton gravity. For instance, asymptotically flat black hole
solutions in the absence of the dilaton potential are studied in Refs. \cite%
{CDB1,CDB2,MW,PW}. In recent years, AdS/CFT correspondence have made
asymptotically non-flat solutions more valuable. On the other hand, study of
asymptotically non-flat and non-AdS solutions extends the validity range of
tools already applied and tested in the cases of asymptotically flat or AdS
spacetimes. Many efforts have been made to find and study static and
rotating nonflat exact solutions in the literature. Static asymptotically
nonflat and non AdS solutions and their thermodynamics were explored in \cite%
{MW,PW,CHM,Cai,Clem,yaz,SRM,Shey2}. Different aspects of exact rotating
nonflat solutions have also been studied in \cite%
{Fr,Dias,Har,Mai,Deh1,Mitra,Shey0,SDR,sheyhendi,Shey3}.

In the present paper we extend the study on the dilaton gravity to include
the power-law Maxwell field term in the action. The motivation for this
study comes from the fact that, as in the case of scalar field which has
been shown that particular power of the massless Klein-Gordon Lagrangian
shows conformal invariance in arbitrary dimensions \cite{Haissene}, one can
have a conformally electrodynamic Lagrangian in higher dimensions. It is
worth mentioning that Maxwell Lagrangian, $F_{\mu \nu }F^{\mu \nu }$ is
conformally invariant only in four dimensions, while it was shown that the
Lagrangian $(F_{\mu \nu }F^{\mu \nu })^{(n+1)/4}$ is conformally invariant
in $(n+1)$-dimensions \cite{Haissene}. In other words, this Lagrangian is
invariant under the conformal transformation $g_{\mu \nu }\rightarrow \Omega
^{2}g_{\mu \nu }$ and $A_{\mu }\rightarrow A_{\mu }$. The studies on the
black object solutions coupled to a conformally invariant Maxwell field have
got a lot of attentions in the past decades \cite%
{Hass1,hendisedehi,confshey2,sad,KZSD}. Thermodynamics of higher dimensional
Ricci flat rotating black branes with a conformally invariant power-Maxwell
source in the absence of a dilaton field were studied in \cite{hendisedehi}.
Recently, we explored exact topological black hole solutions in the presence
of nonlinear power-law Maxwell source as well as dilaton field \cite{KZSD}.
Since these solutions (\cite{KZSD}) are static, it is worthwhile to
construct the rotating version of the solutions. Charged rotating dilaton
black branes in dilaton gravity and their thermodynamics in the presence of
linear Maxwell field and nonlinear Born-Infeld electrodynamics have been
studied in \cite{Shey3} and \cite{DHSR} respectively. Till now, charged
rotating dilaton black objects coupled to a power-law Maxwell field has not
been constructed. In this paper, we intend to construct a new class of $%
(n+1) $-dimensional rotating dilaton black branes in the presence of
power-law Maxwell field, where we relax the conformally invariant issue for
generality. Of course, the solution do exist for the case of conformally
invariant source. We find that the solutions exist provided one takes
Liouville-type potentials with two terms. In the limiting case of Maxwell
field, one of the Liouville potentials vanishes. We shall investigate
thermodynamics as well as thermal stability of our rotating black brane
solutions and explore the effects of the dilaton and power-law Maxwell
fields on thermodynamics and thermal stability of these black branes.

The outline of this paper is as follows. In the next section, we present the
basic field equations of Einstein-dilaton gravity with power-law Maxwell
field and find a new class of charged rotating black brane solutions of this
theory and investigate their properties. In section \ref{Therm}, we obtain
the conserved and thermodynamic quantities of the solutions and verify the
validity of the first law of black hole thermodynamics. In Sec. \ref{Stab},
we study thermal stability of the solutions in both canonical and
grand-canonical ensembles. We finish our paper with closing remarks in the
last section.

\section{FIELD EQUATIONS AND ROTATING SOLUTIONS\label{Field}}

We consider an $(n+1)$-dimensional ($n\geq 3$) action of Einstein gravity
which is coupled to dilaton and power-law Maxwell field, 
\begin{equation}
I=-\frac{1}{16\pi }\int_{\mathcal{M}}d^{n+1}x\sqrt{-g}\left\{ \mathcal{R}%
\text{ }-\frac{4}{n-1}(\nabla \Phi )^{2}-V(\Phi )+\left( -e^{-4\alpha \Phi
/(n-1)}\mathcal{F}\right) ^{p}\right\} -\frac{1}{8\pi }\int_{\partial 
\mathcal{M}}d^{n}x\sqrt{-\gamma }\Theta (\gamma ),  \label{Act}
\end{equation}%
where $\mathcal{R}$ is the Ricci scalar, $\Phi $ is the dilaton field and $%
V(\Phi )$ is the dilaton potential. Here $\mathcal{F}=F^{\mu \nu }F_{\mu \nu
}$ where $F_{\mu \nu }=\partial _{\lbrack \mu }A_{\nu ]}$ is the
electromagnetic field tensor and $A_{\mu }$ is the electromagnetic
potential, while $p$ and $\alpha $ are two constants that determine
nonlinearity of electromagnetic field and coupling strength of the dilaton
and electromagnetic fields, respectively. The well-known Einstein-Maxwell
dilaton theory corresponds to the case $p=1$. The last term in (\ref{Act})
is Gibbons-Hawking boundary term which is added to the action in order to
make variational principle well-defined. We show the metric of manifold $%
\mathcal{M}$ with $g_{\mu \nu }$, with covariant derivative $\nabla _{\mu }$%
. The metric of the boundary $\partial \mathcal{M}$ is $\gamma _{ab}$ and $%
\Theta =$ $\gamma _{ab}\Theta ^{ab}$ is the trace of the extrinsic curvature
of the boundary $\Theta ^{ab}$. By varying action (\ref{Act}) with respect
to the gravitational field $g_{\mu \nu }$, the dilaton field $\Phi $ and the
gauge field $A_{\mu }$, one can obtain equations of motion as 
\begin{eqnarray}
&&\mathcal{R}_{\mu \nu }=g_{\mu \nu }\left\{ \frac{V(\Phi )}{n-1}+\frac{%
(2p-1)}{n-1}\left( -\mathcal{F}e^{-4\alpha \Phi /(n-1)}\right) ^{p}\right\} +%
\frac{4\left( \partial _{\mu }\Phi \partial _{\nu }\Phi \right) }{n-1}+2pe^{-%
{4\alpha p\Phi }/({n-1})}(-\mathcal{F})^{p-1}F_{\mu \lambda }F_{\nu }^{\text{
\ }\lambda },  \label{FE1} \\
&&\nabla ^{2}\Phi =\frac{n-1}{8}\frac{\partial V}{\partial \Phi }+\frac{%
p\alpha }{2}e^{-{4\alpha p\Phi }/({n-1})}(-\mathcal{F})^{p},  \label{FE2} \\
&&\nabla _{\mu }\left( e^{-{4\alpha p\Phi }/({n-1})}(-\mathcal{F}%
)^{p-1}F^{\mu \nu }\right) =0.  \label{FE3}
\end{eqnarray}%
Here we seek for higher dimensional rotating solutions of field equations (%
\ref{FE1})-(\ref{FE3}). We know that rotation group in $(n+1)$-dimensions is 
$SO(n)$. Thus the number of independent rotation parameters for a localized
object is $k\equiv \lbrack n/2]$ where $[x]$ is the integer part of $x$.
Note that $k$ is equal to number of Casimir operators. Therefore, the metric
of $(n+1)$-dimensional rotating solution with cylindrical or toroidal
horizons and complete set of rotation parameters can be written as \cite%
{awad} 
\begin{eqnarray}
ds^{2} &=&-f(r)\left( \Xi dt-{{\sum_{i=1}^{k}}}a_{i}d\phi _{i}\right) ^{2}+%
\frac{r^{2}}{l^{4}}R^{2}(r){{\sum_{i=1}^{k}}}\left( a_{i}dt-\Xi l^{2}d\phi
_{i}\right) ^{2}  \notag \\
&&-\frac{r^{2}}{l^{2}}R^{2}(r){\sum_{i<j}^{k}}(a_{i}d\phi _{j}-a_{j}d\phi
_{i})^{2}+\frac{dr^{2}}{f(r)}+\frac{r^{2}}{l^{2}}R^{2}(r)dX^{2},  \notag \\
\Xi ^{2} &=&1+\sum_{i=1}^{k}\frac{a_{i}^{2}}{l^{2}},  \label{Met3}
\end{eqnarray}%
where $a_{i}$s are $k$ rotation parameters and $l$ has the dimension of
length which is related to the cosmological constant $\Lambda $ for the case
of Liouville-type potential with constant $\Phi $. The angular coordinates
are in the range $0\leq \phi _{i}\leq 2\pi $ and $dX^{2}$ is the Euclidean
metric on the $(n-k-1)$-dimensional submanifold with volume $\Sigma _{n-k-1}$%
. Integrating the Maxwell equation (\ref{FE3}) leads to 
\begin{equation}
F_{tr}=\frac{q\Xi e{^{{\frac{4\alpha \,p\Phi \left( r\right) }{\left(
n-1\right) \left( 2\,p-1\right) }}}}}{\left( rR\right) ^{{\frac{n-1}{2\,p-1}}%
}},\text{ \ \ \ \ \ \ }F_{\phi _{i}r}=-\frac{a_{i}}{\Xi }F_{tr}.  \label{FF}
\end{equation}%
where $q$ is an integration constant related to the electric charge of the
brane. Substituting (\ref{Met3}) and (\ref{FF}) in the field equations (\ref%
{FE1}) and (\ref{FE2}), we find the following differential equations 
\begin{eqnarray}
&&-\frac{{\Xi }^{2}f^{\prime \prime }}{2}-\frac{\left( n-1\right) }{2}\left( 
{\frac{f^{\prime }R^{\prime }}{R}}+{\frac{f^{\prime }}{r}}\right) +\left( {%
\Xi }^{2}-1\right) \left[ {\frac{2\left( n-1\right) fR^{\prime }}{rR}+}%
\left( n-2\right) \left( {\frac{f}{{r}^{2}}}+{\frac{fR^{\prime 2}}{R^{2}}}%
\right) \right.  \notag  \label{fe1} \\
&&\text{ }\left. -\,\frac{\left( n-3\right) }{2}\left( {\frac{f^{\prime
}R^{\prime }}{R}}+{\frac{f^{\prime }}{r}}\right) +{\frac{fR^{\prime \prime }%
}{R}}\right] -{\frac{V(\Phi )}{n-1}}-\frac{{2}^{p}{q}^{2\,p}e{^{\,{\frac{%
4\alpha \,p\Phi }{\left( n-1\right) \left( 2\,p-1\right) }}}}}{\left(
rR\right) ^{{\frac{2p\left( n-1\right) }{2\,p-1}}}}\left( {\frac{2p-1}{n-1}}%
-p{\Xi }^{2}\right) =0,  \notag \\
&&
\end{eqnarray}%
\begin{eqnarray}
&&-{\frac{f^{\prime \prime }}{2}}-\,{\frac{\left( n-1\right) f^{\prime
}R^{\prime }}{2R}}-\,{\frac{\left( n-1\right) f^{\prime }}{2r}}-{\frac{%
4f\Phi ^{\prime 2}}{n-1}}-{\frac{2\left( n-1\right) fR^{\prime }}{rR}}-{%
\frac{\left( n-1\right) fR^{\prime \prime }}{R}}-{\frac{V(\Phi )}{n-1}} 
\notag \\
&&\text{ \ \ \ \ \ }\left. +\frac{{2}^{p}{q}^{2\,p}e{^{{\frac{4\alpha
\,p\Phi }{\left( n-1\right) \left( 2\,p-1\right) }}}}}{\left( rR\right) ^{{%
\frac{2p\left( n-1\right) }{2\,p-1}}}}\left( {\frac{1+p\left( n-3\right) }{%
n-1}}\right) =0,\right.
\end{eqnarray}

\begin{equation}
{\frac{f^{\prime \prime }}{2}}-\,{\frac{2\left( n-1\right) fR^{\prime }}{rR}}%
-{\frac{fR^{\prime \prime }}{R}}+\frac{n-3}{2}\,\left( {\frac{f^{\prime }}{r}%
}+{\frac{R^{\prime }f^{\prime }}{R}}\right) -\left( n-2\right) \left( {\frac{%
f}{{r}^{2}}}+{\frac{fR^{\prime 2}}{R^{2}}}\right) -\frac{{2}^{p}p{q}^{2\,p}e{%
^{{\frac{4\alpha \,p\Phi }{\left( n-1\right) \left( 2\,p-1\right) }}}}}{%
\left( rR\right) ^{{\frac{2p\left( n-1\right) }{2\,p-1}}}}=0,
\end{equation}

\begin{eqnarray}
&&{\frac{a_{i}^{2}f^{\prime \prime }}{2{l}^{2}}+}\left( 1+{\frac{a_{i}^{2}}{{%
l}^{2}}}\right) \left[ -2\,\left( n-1\right) \frac{fR^{\prime }}{{r}R}-\frac{%
fR^{\prime \prime }}{R}-\left( n-2\right) \left( \frac{f}{{r}^{2}}+\frac{%
fR^{\prime 2}}{R^{2}}\right) \right]  \notag \\
&&\text{ \ \ \ \ \ \ \ }+\left( \frac{\,\left( n-3\right) a_{i}^{2}}{2{l}^{2}%
}-1\right) \left( {\frac{f^{\prime }}{r}}+{\frac{R^{\prime }f^{\prime }}{R}}%
\right) -{\frac{V(\Phi )}{n-1}}-\frac{{2}^{p}{q}^{2\,p}e{^{{\frac{4\alpha
\,p\Phi }{\left( n-1\right) \left( 2\,p-1\right) }}}}}{\left( rR\right) ^{{%
\frac{2p\left( n-1\right) }{2\,p-1}}}}\left( {\frac{2p-1}{n-1}}+{\frac{%
pa_{i}^{2}}{{l}^{2}}}\right) =0,  \notag \\
&&
\end{eqnarray}%
\begin{equation}
-\left( n-2\right) f\left( \frac{1}{{r}^{2}}+{\frac{R^{\prime 2}}{R^{2}}}%
\right) -{\frac{2\left( n-1\right) fR^{\prime }}{rR}}-{\frac{f^{\prime }}{r}}%
-{\frac{R^{\prime }f^{\prime }}{R}}-{\frac{fR^{\prime \prime }}{R}}-{\frac{%
V(\Phi )}{n-1}}-\frac{{2}^{p}{q}^{2\,p}e{^{{\frac{4\alpha \,p\Phi }{\left(
n-1\right) \left( 2\,p-1\right) }}}}}{\left( rR\right) ^{{\frac{2p\left(
n-1\right) }{2\,p-1}}}}\left( {\frac{2p-1}{n-1}}\right) =0,
\end{equation}%
\begin{equation}
f\Phi ^{\prime \prime }+\left( n-1\right) f\Phi ^{\prime }\left( \frac{1}{r}+%
{\frac{R^{\prime }}{R}}\right) +f^{\prime }\Phi ^{\prime }-\frac{\left(
n-1\right) }{8}\frac{dV(\Phi )}{d\Phi }-\frac{{2}^{p-1}p\alpha \,{q}^{2\,p}e{%
^{{\frac{4\alpha \,p\Phi }{\left( n-1\right) \left( 2\,p-1\right) }}}}}{%
\left( rR\right) ^{{\frac{2p\left( n-1\right) }{2\,p-1}}}}=0,
\end{equation}%
\begin{equation}
{\frac{f^{\prime \prime }}{2}}-\left( n-2\right) f\left( \frac{1}{{r}^{2}}+{%
\frac{R^{\prime 2}}{R^{2}}}\right) +\frac{\,\left( n-3\right) }{2}\left( {%
\frac{f^{\prime }}{r}}+{\frac{R^{\prime }f^{\prime }}{R}}\right) -{\frac{%
fR^{\prime \prime }}{R}}-{\frac{2\left( n-1\right) fR^{\prime }}{rR}}-\frac{{%
2}^{p}p{q}^{2\,p}e{^{{\frac{4\alpha \,p\Phi }{\left( n-1\right) \left(
2\,p-1\right) }}}}}{\left( rR\right) ^{{\frac{2p\left( n-1\right) }{2\,p-1}}}%
}=0,  \label{fe7}
\end{equation}%
where the prime denotes derivative with respect to $r$. In order to
construct exact rotating solutions of the theory given by action (\ref{Act}%
), the functions $f(r)$, $R(r)$ and $\Phi (r)$ should be determined so that
the system of equations (\ref{fe1})-(\ref{fe7}) are satisfied. To do that,
we make the ansatz \cite{Shey3}%
\begin{equation}
R(r)=e^{2\alpha \Phi (r)/(n-1)},  \label{Rphi}
\end{equation}%
and take the potential of the Liouville-type with two terms, namely 
\begin{equation}
V(\Phi )=2\,\Lambda _{1}e{^{2\zeta _{1}\,\Phi }}+2\,\Lambda e{^{2\zeta
_{2}\,\Phi }},  \label{v1}
\end{equation}%
where $\Lambda _{1}$, $\Lambda $, $\zeta _{1}$ and $\zeta _{2}$ are
constants. Using (\ref{Rphi}) and (\ref{v1}), one can easily show that
equations (\ref{fe1})-(\ref{fe7}) have solutions of the form 
\begin{equation}
f(r)=-\frac{2\Lambda {b}^{2\gamma }(1+\alpha ^{2})^{2}{r}^{2(1-\gamma )}}{%
\left( n-1\right) \left( n-{\alpha }^{2}\right) }-\frac{m}{{r}%
^{(n-1)(1-\gamma )-1}}+\frac{2^{p}p(1+\alpha ^{2})\left( 2p-1\right) {q}%
^{2\,p}\,}{\Pi \Upsilon {b}^{{2\left( n-2\right) p\gamma /}\left(
2\,p-1\right) }{r}^{\Upsilon +(n-1)(1-\gamma )-1}},  \label{fr}
\end{equation}

\begin{equation}
\Phi (r)=\frac{(n-1)\alpha }{2(1+\alpha ^{2})}\ln \left( \frac{b}{r}\right) ,
\label{phi}
\end{equation}%
where $b$\ and $m$\ are arbitrary constants, $\gamma =\alpha ^{2}/(\alpha
^{2}+1)$, $\Upsilon =(n-2p+\alpha ^{2})/(2p-1)(1+\alpha ^{2})$ and $\Pi
=\alpha ^{2}+\left( n-1-{\alpha }^{2}\right) p$. The above solutions will
fully satisfy the system of equations provided, 
\begin{equation*}
\zeta _{1}={\frac{2p\left( n-1+{\alpha }^{2}\right) }{\left( n-1\right)
\left( 2\,p-1\right) \alpha }},\text{ \ \ \ \ }\zeta _{2}=\,{\frac{2\alpha }{%
n-1}},\text{ \ \ \ \ }\Lambda _{1}=\frac{2^{p-1}\left( 2\,p-1\right) \left(
p-1\right) {\alpha }^{2}\,{q}^{2\,p}}{\Pi {b}^{{\frac{2\left( n-1\right) p}{%
2\,p-1}}}}.
\end{equation*}%
It is worthwhile to note that the necessity of existence of the term $%
2\Lambda _{1}e^{2\zeta _{1}\,\Phi }$ in scalar field potential is due to
nonlinearity of Maxwell field and this term vanishes for the case of linear
Maxwell field where $p=1$ \cite{Shey3}. Note that in our solutions $\Lambda $%
\ remains as a free parameter which plays the role of the cosmological
constant. Another thing to notice is that although these rotating solutions
are locally the same as those found in \cite{KZSD} with flat horizon ($k=0$%
), they are not the same globally. One may also note that in the particular
case $p=1$\ these solutions reduce to the $(n+1)$-dimensional charged
rotating dilaton black branes presented in \cite{Shey3}. The parameter $m$\
in Eq. (\ref{fr}) is the integration constant which is known as the
geometrical mass and can be written in term of the horizon radius as%
\begin{equation}
m(r_{+})=\frac{2^{p}p\,\left( 2p-1\right) {q}^{2\,p}}{(1-\gamma )\Upsilon
\Pi {b}^{{2\left( n-2\right) \gamma p/}\left( 2\,p-1\right) }{r}%
_{+}^{\Upsilon }}\,+\frac{{b}^{2\gamma \,}n{r}_{+}^{(1-\gamma )({n+1)-1}}}{%
l^{2}(1-\gamma )^{2}\left( n-{\alpha }^{2}\right) },  \label{mrh}
\end{equation}%
where $r_{+}$\ is the positive real root of $f(r_{+})=0$. One can easily
show that the vector potential $A_{\mu }$\ corresponding to the
electromagnetic tensor (\ref{FF}) can be written as%
\begin{equation}
A_{\mu }=\frac{q{b}^{{\frac{\left( 2\,p+1-n\right) \gamma }{\left(
2\,p-1\right) }}}}{\Upsilon {r}^{\Upsilon }}\left( \Xi \delta _{\mu
}^{t}-a_{i}\delta _{\mu }^{i}\right) \hspace{0.5cm}{\text{(no sum on }i\text{%
)}}.
\end{equation}%
Here we pause to give some remarks about the value of $A_{\mu }$\ at
infinity that apply some restrictions on the value of $p$ and $\alpha $.
This discussion is given here because it is necessary to know these
restrictions for next studies. In section (\ref{Therm}), we will show that
the mass of solutions is dependent on the value of $A_{\mu }$\ at infinity.
Thus, the value of electromagnetic vector potential should be finite at
infinity so that we have finite mass. In order to guarantee this behavior $%
\Upsilon $ should be positive i.e.%
\begin{equation}
\frac{{n-2p+}\alpha ^{2}}{({2p-1)(1+}\alpha ^{2})}>0.  \label{res2}
\end{equation}%
The above equation leads to the following restriction on the range of $p$, 
\begin{equation}
\frac{1}{2}<p<\frac{n+\alpha ^{2}}{2}.  \label{res3}
\end{equation}%
On the other hand, the effect of $m$\ in metric function should vanish in
spacial infinity. This fact leads to a restriction on $\alpha $\ 
\begin{equation}
\alpha ^{2}<n-2.  \label{res5}
\end{equation}%
Therefore, one can sum up the constraints (\ref{res3}) and (\ref{res5}) as
follows:

\begin{eqnarray}
\text{for }\frac{1}{2} &<&p<\frac{n}{2}\text{, \ \ \ \ \ \ \ \ }0\leq \alpha
^{2}<n-2,  \label{res6} \\
\text{for }\frac{n}{2} &<&p<n-1\text{, \ \ \ }2p-n<\alpha ^{2}<n-2.
\label{res7}
\end{eqnarray}%
It is worth mentioning that the solution is always well-defined in the above
allowed ranges of $\alpha $\ and $p$. In continue of this section we discuss
properties and asymptotic behaviors of the solutions in allowed ranges of $p$
and $\alpha $.

\subsection{Asymptotic behaviors and properties of the solutions}

Now, by considering the restrictions (\ref{res6}) and (\ref{res7}) on $p$
and $\alpha $, we are ready to discuss the behavior of our solutions both in
the vicinity of $r=0$ and infinity. First, one should note that in allowed
ranges of $p$ and $\alpha $, the charge term in metric function disappears
at infinity as the mass term does. This behavior is also seen in the special
cases of $p=1$\ or $\alpha =0$. Therefore, the first term in $f(r)$ ((\ref%
{fr})) determines the behavior of it at infinity. Since we are interested in
the solutions that go to infinity as $r\rightarrow \infty $, we assume $%
\Lambda <0$ and take it in the standard form $\Lambda =-n(n-1)/2l^{2}$. It
is also notable to mention that the spacetime is neither asymptotically flat
nor AdS due to existence of dilaton field. About $r=0$, the term including $%
q $ in metric function is dominant. This term is always positive in
permitted ranges of $p$ and $\alpha $ because $\Pi ,\Upsilon >0$ in these
ranges. Therefore as in the case of Reissner-Nordstrom black holes, we have
timelike singularity and there are no Schwarzschild-type black hole
solutions.

\begin{figure}[h]
\epsfxsize=7cm\centerline{\epsffile{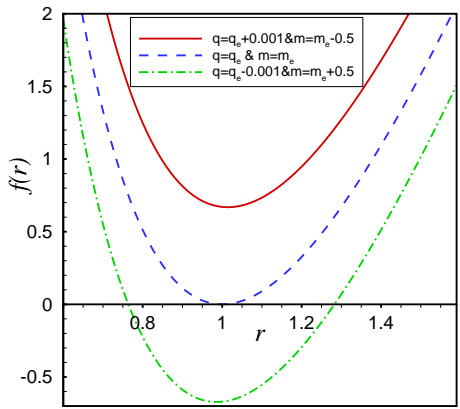}}
\caption{The behavior of $f(r)$ versus $r$ with $l=b=1$, $q=0.5$, $\protect%
\alpha =\protect\sqrt{2}$, $n=5$, $p=2$, $r_{\mathrm{ext}}=1$. In this case $%
m_{\mathrm{ext}}=60.00$ and $q_{\mathrm{ext}}=1.06$.}
\label{fig1}
\end{figure}

We continue our discussions about properties of the solutions by looking for
curvature singularities. Since essential singularities are located at
divergencies of Kretschmann scalar, we seek for these for our solutions. It
is easy to show that the Kretschmann scalar $R_{\mu \nu \lambda \kappa
}R^{\mu \nu \lambda \kappa }$ diverges at $r=0$ while it is finite for $%
r\neq 0$\ and goes to zero as $r\rightarrow \infty $ and therefore there is
an essential singularity located at $r=0$. Although the location of event
horizon cannot be determined analytically by using $f(r)$, fortunately we
can get more insight about the solutions by calculating temperature
corresponding to event horizon. The temperature and angular velocity of the
horizon can be obtained by analytic continuation of the metric. The
analytical continuation of the Lorentzian metric by $t\rightarrow i\tau $\
and $a\rightarrow ia$\ yields the Euclidean section. In order that Euclidean
metric is regular at $r=r_{+}$, one should identify $\tau \sim \tau +\beta
_{+}$\ and $\phi _{i}\sim \phi _{i}+\beta _{+}\Omega _{i}$, where $\beta
_{+} $\ and $\Omega _{i}$s are the inverse Hawking temperature and the $i$th
component of angular velocity of the horizon. Then, temperature and $i$th
component of angular velocity can be computed as

\begin{eqnarray}
T_{+} &=&\frac{f^{\prime }(r_{+})}{4\pi \Xi }=\frac{(1+\alpha ^{2})}{4\pi
\Xi }\left\{ \frac{nb^{2\gamma }r_{+}^{1-2\gamma }}{l^{2}}-\frac{%
2^{p}p\left( 2p-1\right) {q}^{2\,p}}{\Pi {b}^{\,{2\left( n-2\right) \gamma p/%
}\left( 2\,p-1\right) }{r}_{+}^{\Upsilon +(1-\gamma )\left( n-1\right) }}%
\right\} ,  \label{Tem} \\
\Omega _{i} &=&\frac{a_{i}}{\Xi l^{2}}.  \label{Om1}
\end{eqnarray}%
Numerical calculations show that temperature vanishes at event horizon for
extreme black brane solutions. Therefore, one can see from Eq. (\ref{Tem})
that we have extreme black brane if 
\begin{equation}
m_{\mathrm{ext}}=\frac{(\alpha ^{2}+1)n{b}^{2\gamma \,}}{\Upsilon l^{2}}%
\left[ 1+\frac{(1+\alpha ^{2})\Upsilon }{\left( n-{\alpha }^{2}\right) }%
\right] \,{r}_{\mathrm{ext}}^{{(1-\gamma )}(n+1{)-1}},
\end{equation}%
or%
\begin{equation}
{q}_{\mathrm{ext}}^{2p}=\frac{\Pi nb^{2\gamma \left( np-1\right) /(2p-1)}}{%
2^{p}p\left( 2p-1\right) l^{2}}{r}_{\mathrm{ext}}^{\Upsilon +(1-\gamma
)(n+1)-1}.
\end{equation}%
The solutions have two inner and outer horizons located at $r_{-}$ and $%
r_{+} $, provided the charge parameter $q$ is lower than $q_{\mathrm{ext}}$
or $m$ is greater than $m_{\mathrm{ext}}$ and a naked singularity if $q>q_{%
\mathrm{ext}}$ or $m<m_{\mathrm{ext}}$ (see Fig. \ref{fig1}). Note that
there is a relation between $m_{\mathrm{ext}}$ and $q_{\mathrm{ext}}$ as

\begin{equation}
m_{\mathrm{ext}}=\left( \frac{(\alpha ^{2}+1)n{b}^{2\gamma \,}\left( \left(
n-{\alpha }^{2}\right) +(1+\alpha ^{2})\Upsilon \right) }{\Upsilon
l^{2}\left( n-{\alpha }^{2}\right) }\right) \left( \frac{2^{p}p\left(
2p-1\right) l^{2}}{\Pi nb^{2\gamma \left( np-1\right) /(2p-1)}}{q}_{\mathrm{%
ext}}^{2p}\right) ^{\frac{{(1-\gamma )}(n+1{)-1}}{\Upsilon +(1-\gamma
)(n+1)-1}},  \label{mext}
\end{equation}%
which reduces to extremal mass obtained in \cite{Shey3} for linear Maxwell
field ($p=1$) and to one obtained in \cite{Deh3} in the absence of dilaton
field ${\alpha =}\gamma =0$ and with linear Maxwell field. Finally, it is
noticeable to mention that there is a Killing horizon in addition to event
horizon for rotating solutions in our Einstein-dilaton gravity as in the
case of rotating black solutions of the Einstein gravity. It is easy to show
that the Killing vector field%
\begin{gather}
\tilde{\chi}=C\chi ,  \notag \\
\chi =\partial _{t}+{{{\sum_{i=1}^{k}}}}\Omega _{i}\partial _{\phi _{i}},
\label{chi}
\end{gather}%
is the null generator of the event horizon, where $k$\ denote the number of
rotation parameters \cite{Deh4} and $C$\ is a constant that we will fix it
in next section. The Killing horizon is a null surface whose null generators
are tangent to a Killing field. 

\section{THERMODYNAMICS OF BLACK BRANES}

\label{Therm}In this section, we discuss thermodynamics of rotating black
brane solutions in the presence of power-law Maxwell field. Since the
discussion of thermodynamics of these solutions depends on the calculation
of the mass and other conserved charges of the spacetime, we first compute
these conserved quantities. The way we use for calculating conserved
quantities is counterterm method. This method is a well-known method for
Asymptotically Ads solutions to avoid divergencies in calculation of
conserved quantities which is inspired by AdS/CFT correspondence \cite{Mal}.
This method can also be used for dilaton gravity and in the presence of
Liouvill-type potential where the spacetime is not asymptotically AdS \cite%
{Deh1,shads,Shey3}. Since boundary curvature of our spacetime is zero ($%
R_{abcd}(\gamma )=0$), the counterterm for the stress energy tensor should
be proportional to $\gamma ^{ab}$. We find the finite stress-energy tensor
in $(n+1)$-dimensional Einstein-dilaton gravity with Liouville-type
potential as 
\begin{equation}
T^{ab}=\frac{1}{8\pi }\left[ \Theta ^{ab}-\Theta \gamma ^{ab}+\frac{n-1}{l_{%
\mathrm{eff}}}\gamma ^{ab}\right] ,  \label{Stres}
\end{equation}%
where $l_{\mathrm{eff}}$ is given by%
\begin{equation}
l_{\mathrm{eff}}^{2}=\frac{(n-1)(\alpha ^{2}-n)}{V(\Phi )}.  \label{leff}
\end{equation}%
The first two terms in (\ref{Stres}) are the variation of the action (\ref%
{Act}) with respect to $\gamma _{ab}$, and the last term is counterterm
which removes the divergences. Note that in the absence of the dilaton field
($\alpha =0$), we have $V(\Phi )=2\Lambda $, and the effective $l_{\mathrm{%
eff}}^{2}$ of Eq. (\ref{leff}) reduces to $l^{2}=-n(n-1)/2\Lambda $ of the
AdS spacetimes. In order to compute the conserved charges of the spacetime,
one should first choose a spacelike surface $\mathcal{B}$\ in $\partial 
\mathcal{M}$\ with metric $\sigma _{ij}$, and write the boundary metric in
ADM (Arnowitt-Deser-Misner) form 
\begin{equation*}
\gamma _{ab}dx^{a}dx^{a}=-N^{2}dt^{2}+\sigma _{ij}\left( d\varphi
^{i}+V^{i}dt\right) \left( d\varphi ^{j}+V^{j}dt\right) ,
\end{equation*}%
where the coordinates $\varphi ^{i}$\ are the angular variables
parameterizing the hypersurface of constant $r$\ around the origin, and $N$\
and $V^{i}$\ are the lapse and shift functions respectively. Then, the
quasilocal conserved quantities associated with the stress tensors of Eq. (%
\ref{Stres}) can be written as 
\begin{equation}
Q(\mathcal{\xi )}=\int_{\mathcal{B}}d^{n-1}\varphi \sqrt{\sigma }T_{ab}n^{a}%
\mathcal{\xi }^{b},  \label{charge}
\end{equation}%
where $\sigma $\ is the determinant of the metric $\sigma _{ij}$, $\mathcal{%
\xi }$\ and $n^{a}$\ are the Killing vector field and the unit normal vector
on the boundary $\mathcal{B}$. In order to calculate quasilocal mass and
angular momentum, one should choose boundaries with timelike ($\xi =\partial
/\partial t$) and rotational ($\varsigma =\partial /\partial \varphi $)
Killing vector fields i.e.%
\begin{equation}
M=\int_{\mathcal{B}}d^{n-1}\varphi \sqrt{\sigma }T_{ab}n^{a}\xi ^{b},
\label{Mastot}
\end{equation}%
and%
\begin{equation}
J=\int_{\mathcal{B}}d^{n-1}\varphi \sqrt{\sigma }T_{ab}n^{a}\varsigma ^{b},
\label{Angtot}
\end{equation}
provided the surface $\mathcal{B}$\ contains the orbits of $\varsigma $. (%
\ref{Mastot}) and (\ref{Angtot}) \ are conserved mass and angular momenta of
the black hole surrounded by the boundary $\mathcal{B}$.\ It is remarkable
to mention that although mass and angular momenta do not depend on specific
choice of foliation $\mathcal{B}$\ within the hypersurface $\partial 
\mathcal{M}$, they are dependent on location of boundary $\mathcal{B}$\ in
the spacetime. Taking into account the cylindrical symmetry of the rotating
black brane with $k$ rotation parameters along the angular coordinates $%
0\leq \phi _{i}\leq 2\pi $, we denote the volume of the hypersurface
boundary at constant $t$\ and $r$ by $V_{n-1}$. Then, the mass and angular
momentum per unit volume $V_{n-1}$ of the black branes can be calculated
through the use of Eqs. (\ref{Mastot}) and (\ref{Angtot}). We find 
\begin{equation}
{M}=\frac{b^{(n-1)\gamma }}{16\pi l^{n-2}}\left( \frac{(n-\alpha ^{2})\Xi
^{2}+\alpha ^{2}-1}{1+\alpha ^{2}}\right) m,  \label{M}
\end{equation}%
\begin{equation}
J_{i}=\frac{b^{(n-1)\gamma }}{16\pi l^{n-2}}\left( \frac{n-\alpha ^{2}}{%
1+\alpha ^{2}}\right) \Xi ma_{i}.  \label{J}
\end{equation}%
As one can see from (\ref{J}), the angular momentum per unit volume \ is
proportional to $a_{i}$s and therefore it vanishes if $a_{i}=0$\ ($\Xi =1$).
Thus, it is physically reasonable to consider $a_{i}$s as rotational
parameters of the spacetime. The last conserved quantity of our solutions is
electric charge. Electric charge can be obtained by calculating the flux of
the electric field at infinity. By projecting the electromagnetic field
tensor on specific hypersurfaces, we can find the electric field as $E^{\mu
}=g^{\mu \rho }e^{-4\alpha p\Phi /(n-1)}\left( -F\right) ^{p-1}F_{\rho \nu
}u^{\nu }$ where $u^{\nu }$ is normal to such hypersurfaces. The components
of $u^{\nu }$ are%
\begin{equation}
u^{0}=\frac{1}{N},\text{ \ }u^{r}=0,\text{ \ }u^{i}=-\frac{V^{i}}{N},
\end{equation}%
where $N$\ and $V^{i}$\ are the lapse function and shift vector. Eventually,
the electric charge per unit volume $V_{n-1}$ can be calculated as%
\begin{gather}
{Q}=\frac{\Xi \tilde{q}}{4\pi l^{n-2}},  \notag \\
\tilde{q}=\,{2^{p-1}{q}^{2\,p-1}\,.}  \label{chden}
\end{gather}%
One may note that $\tilde{q}=q$\ for $p=1$ \cite{Shey3}.

Now, we calculate thermodynamic quantities entropy $S$ and electric
potential $U$. Entropy of almost all black solutions including ones in
Einstein gravity typically obeys the so called area law \cite{Beck,hunt}.
Dilaton black solutions are not exceptions (see for instance \cite%
{Shey3,KZSD}). Thus, we can calculate the entropy per unit volume $V_{n-1}$\
of our rotating black brane as 
\begin{equation}
{S}=\frac{\Xi b^{(n-1)\gamma }r_{+}^{(n-1)(1-\gamma )}}{4l^{n-2}},
\label{Ent1}
\end{equation}%
The electric potential $U$, can be calculated through the definition \cite%
{Cal}%
\begin{equation}
U=A_{\mu }\tilde{\chi}^{\mu }\left\vert _{r\rightarrow \infty }-A_{\mu }%
\tilde{\chi}^{\mu }\right\vert _{r=r_{+}},  \label{U}
\end{equation}%
where $\tilde{\chi}$\ is the null generators of the event horizon given by
Eq. (\ref{chi}). Therefore, the electric potential may be obtained as%
\begin{equation}
U=\frac{Cq{b}^{{\frac{\left( 2\,p+1-n\right) \gamma }{\left( 2\,p-1\right) }}%
}}{\Xi \Upsilon {r}^{\Upsilon }}.  \label{Pott}
\end{equation}%
Here, we are ready to seek for satisfaction of thermodynamics first law.
First, we should obtain the mass $M$ in terms of extensive quantities $S$, $Q
$\ and $\mathbf{J}$. Using Eqs. (\ref{M})-(\ref{chden}) and the fact that $%
f(r_{+})=0$, we receive%
\begin{equation}
M(S,Q,\mathbf{J})=\frac{\left( (n-\alpha ^{2})Z+\alpha ^{2}-1\right) \mathbf{%
J}}{(n-\alpha ^{2})l\sqrt{Z(Z-1)}},  \label{Msmarr}
\end{equation}%
where $\mathbf{J=}\sqrt{\sum_{i}^{k}{J_{i}}^{2}}$ and $Z=\Xi ^{2}$\ which is
the positive real root of the following equation:

\begin{eqnarray}
&&\left( {\alpha }^{2}+1\right) \sqrt{Z-1}\left( n-{\alpha }^{2}\right)
p\left( 2p-1\right) ^{2}\left[ \frac{\left( 3{2\pi }^{2}Q^{2}\right) ^{{p}}{l%
}^{n+4(p-1)}{b}^{{{\alpha }^{2}}}}{\sqrt{Z}^{\left( \alpha
^{2}+2n-2p-1\right) /\left( n-1\right) }}\left( \frac{4l^{n-2}{S}}{Z}\right)
^{(n-2\,p+{\alpha }^{2})/\left( 1-n\right) }\right] ^{1/(2p-1)}  \notag \\
&&\left. +\left( n-2p+\alpha ^{2}\right) \Pi \left[ -\frac{\left( {\alpha }%
^{2}+1\right) \sqrt{Z-1}{b}^{{\alpha }^{2}}n}{{l}^{n-2}\sqrt{Z}^{{({\alpha }%
^{2}-2n+1)/(n-1)}}}\left( \frac{4l^{n-2}{S}}{Z}\right) ^{{(n-{\alpha }%
^{2})/(n-1)}}+16\,\pi \,l\mathbf{J}\right] =0.\right.   \notag \\
&&
\end{eqnarray}%
Considering $S$, $Q$\ and $\mathbf{J}$\ as a complete set of extensive
quantities for the mass $M(S,Q,\mathbf{J})$, we should define conjugate
intensive quantities to them. These quantities are temperature, angular
velocities and electric potential%
\begin{equation}
T=\left( \frac{\partial {M}}{\partial {S}}\right) _{\mathbf{J},Q},\Omega
_{i}=\left( \frac{\partial {M}}{\partial {J_{i}}}\right) _{S,Q},U=\left( 
\frac{\partial {M}}{\partial {Q}}\right) _{S,\mathbf{J}}.  \label{inte}
\end{equation}%
One can check numerically that quantities defined by Eq. (\ref{inte})
coincide with Eqs. (\ref{Tem}), (\ref{Om1}) and (\ref{Pott}) provided $C$\
is chosen as $C=\left( n-1\right) p^{2}/\Pi $. Note that in the case of
linear Maxwell field ($p=1$), $C$ reduces to $1$ as we expect \cite{Shey3}.
Therefore one can conclude that thermodynamics first law%
\begin{equation}
dM=TdS+{{{\sum_{i=1}^{k}}}}\Omega _{i}d{J}_{i}+Ud{Q},
\end{equation}%
is satisfied. As one can see from (\ref{U}), $U$ is proportional to value of 
$A_{\mu }$ at infinity and therefore $U$ diverges if $A_{\mu }$ diverges at
infinity. On the other hand, it is obvious from thermodynamics first law
that $M$ is dependent on the value of $U$. Therefore, as we mentioned in
section (\ref{Field}), $A_{\mu }$ should be finite at infinity in order to
have a finite mass. We took this fact into account for finding constraints
on $p$ and $\alpha $ (see Eq. (\ref{res2})).

\section{STABILITY IN CANONICAL AND GRAND-CANONICAL ENSEMBLES\label{Stab}}

\begin{figure}[h]
\epsfxsize=7cm \centerline{\epsffile{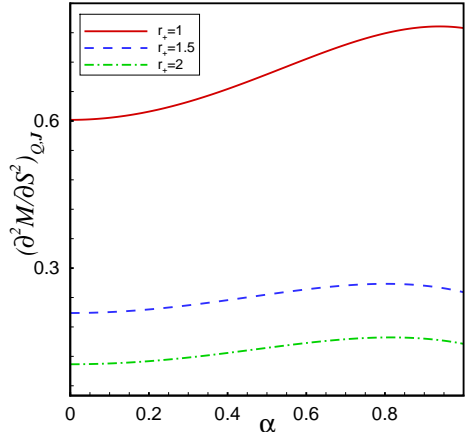}}
\caption{The behavior of $(\partial ^{2}M/\partial S^{2})_{Q,\mathbf{J}}$
versus $\protect\alpha \leq 1$ with $l=b=1$, $q=0.8$, $\Xi =1.25$, $n=5$ and 
$p=2$. }
\label{fig2}
\end{figure}

\begin{figure}[h]
\epsfxsize=7cm \centerline{\epsffile{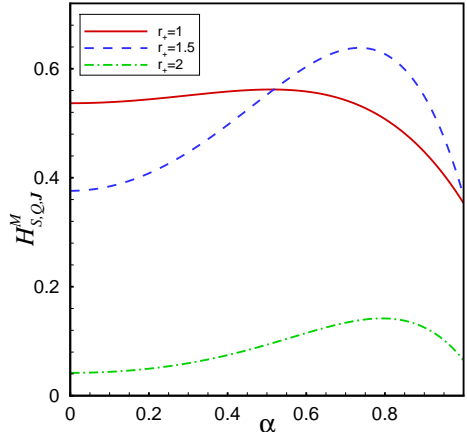}}
\caption{The behavior of $\mathbf{H}_{SQ\mathbf{J}}^{M}$ versus $\protect%
\alpha \leq 1$ with $l=b=1$, $q=0.8$, $\Xi =1.25$, $n=5$ and $p=2$. Note
that curve corresponding to $r_{+}=1$ has been devided by $10$.}
\label{fig3}
\end{figure}

\begin{figure}[h]
\epsfxsize=7cm \centerline{\epsffile{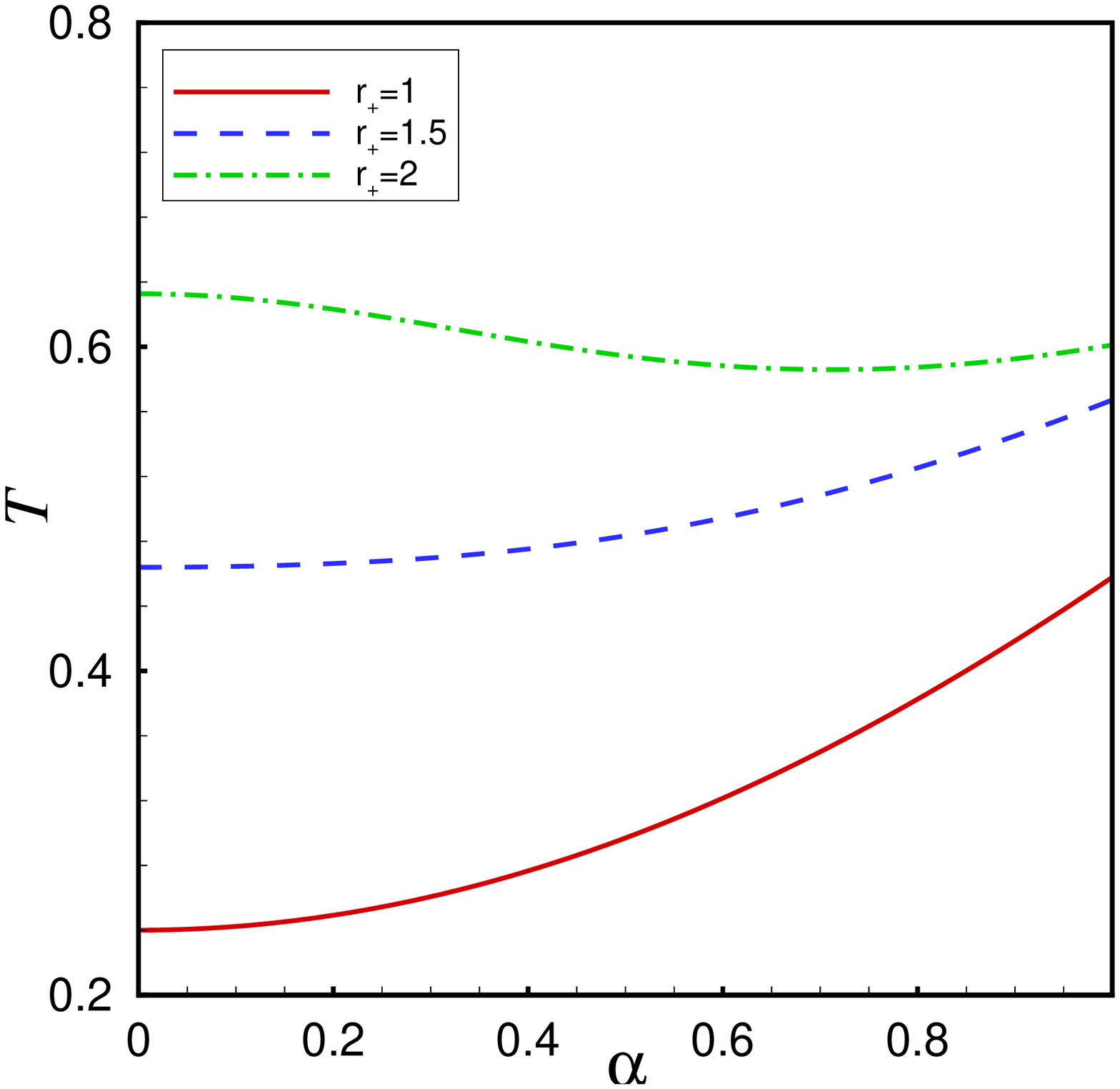}}
\caption{The behavior of $T$ versus $\protect\alpha \leq 1$ with $l=b=1$, $%
q=0.8$, $\Xi =1.25$, $n=5$ and $p=2$.}
\label{fig4}
\end{figure}

\begin{figure}[h]
\epsfxsize=7cm \centerline{\epsffile{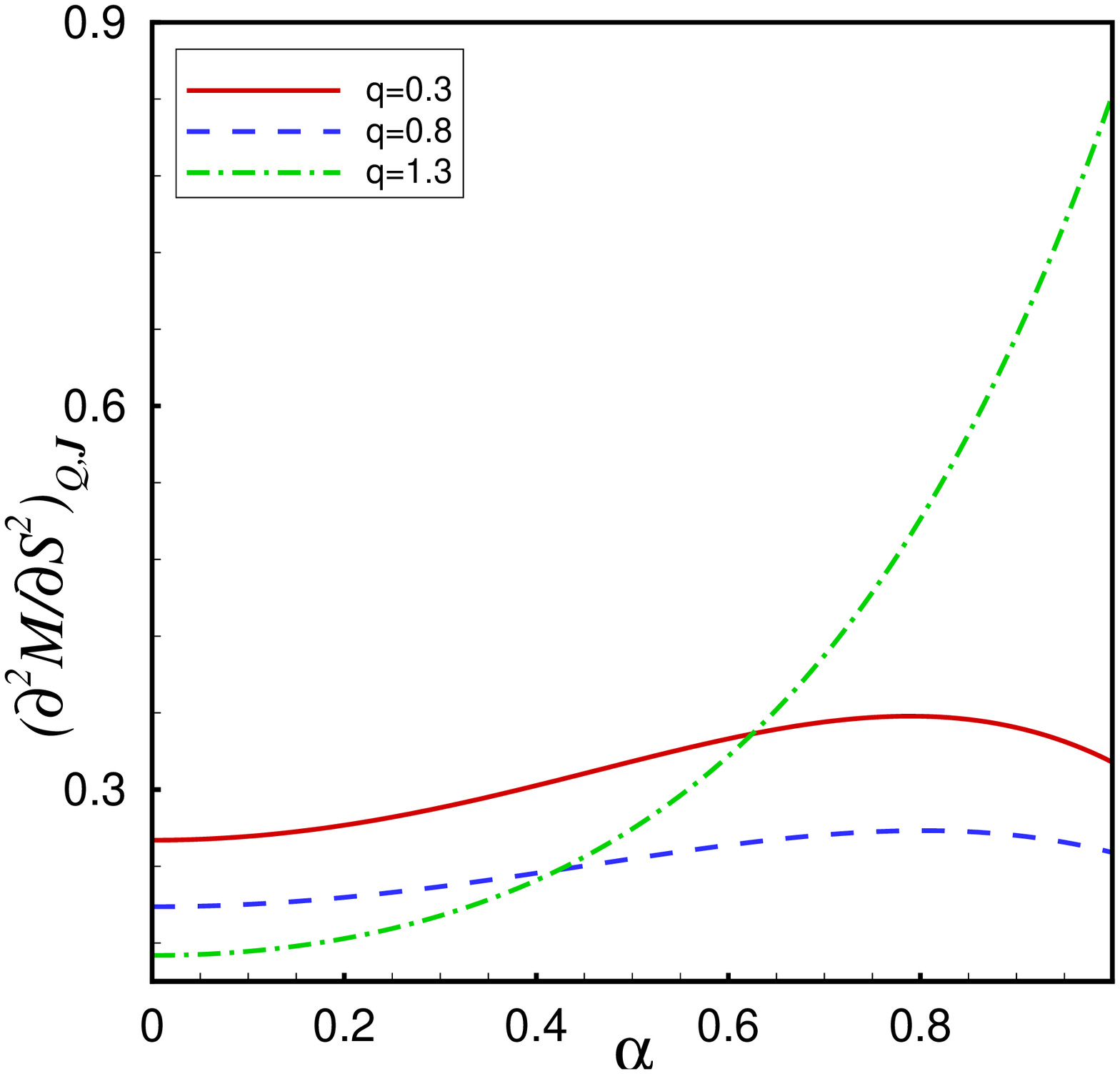}}
\caption{The behavior of $(\partial ^{2}M/\partial S^{2})_{Q,\mathbf{J}}$
versus $\protect\alpha \leq 1$ with $l=b=1$, $r_{+}=1.5$, $\Xi =1.25$, $n=5$
and $p=2$. }
\label{fig5}
\end{figure}

\begin{figure}[h]
\epsfxsize=7cm \centerline{\epsffile{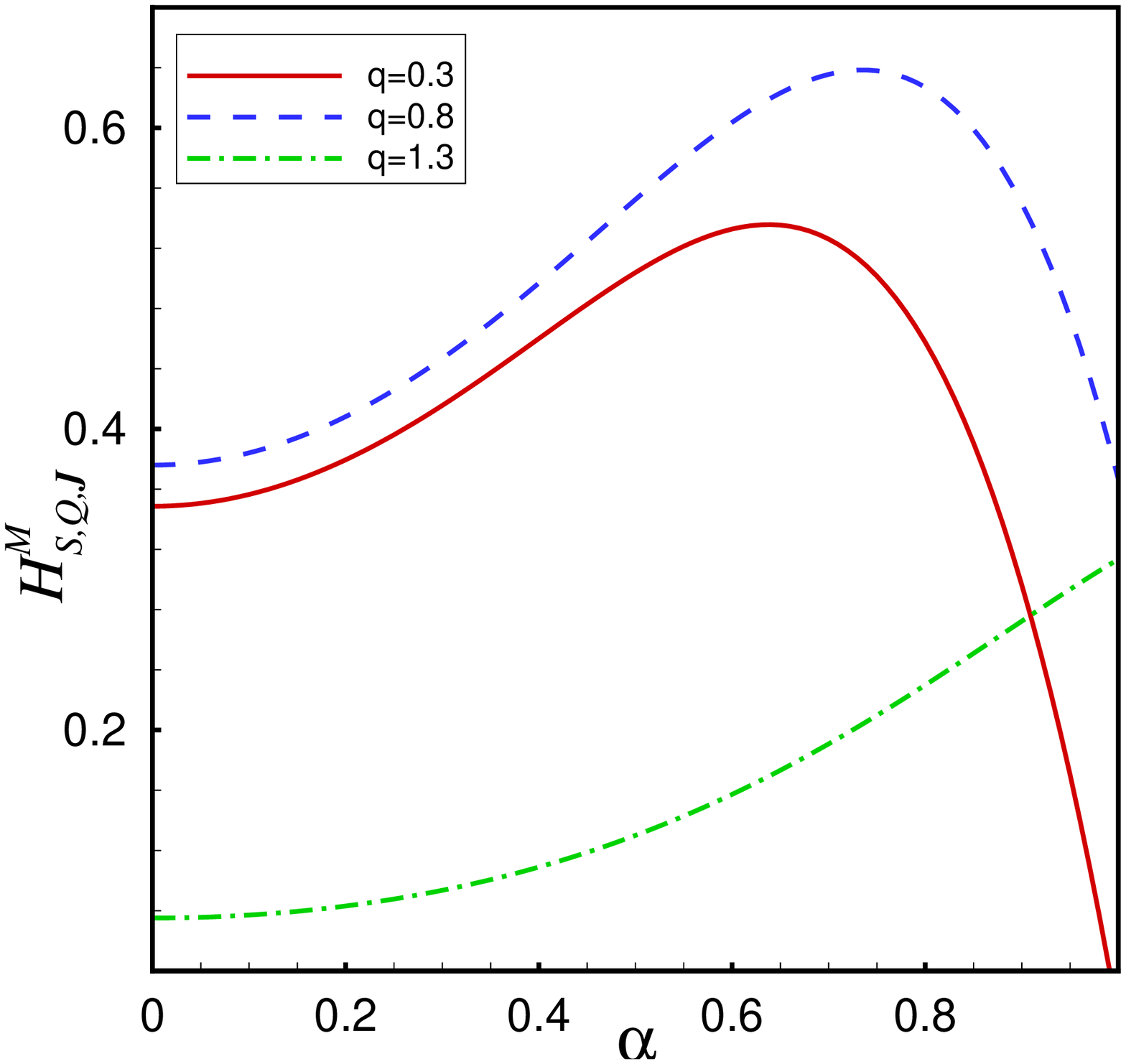}}
\caption{The behavior of $\mathbf{H}_{SQ\mathbf{J}}^{M}$ versus $\protect%
\alpha \leq 1$ with $l=b=1$, $r_{+}=1.5$, $\Xi =1.25$, $n=5$ and $p=2$. Note
that curve corresponding to $q=0.3$ has been devided by $10$.}
\label{fig6}
\end{figure}

\begin{figure}[h]
\epsfxsize=7cm \centerline{\epsffile{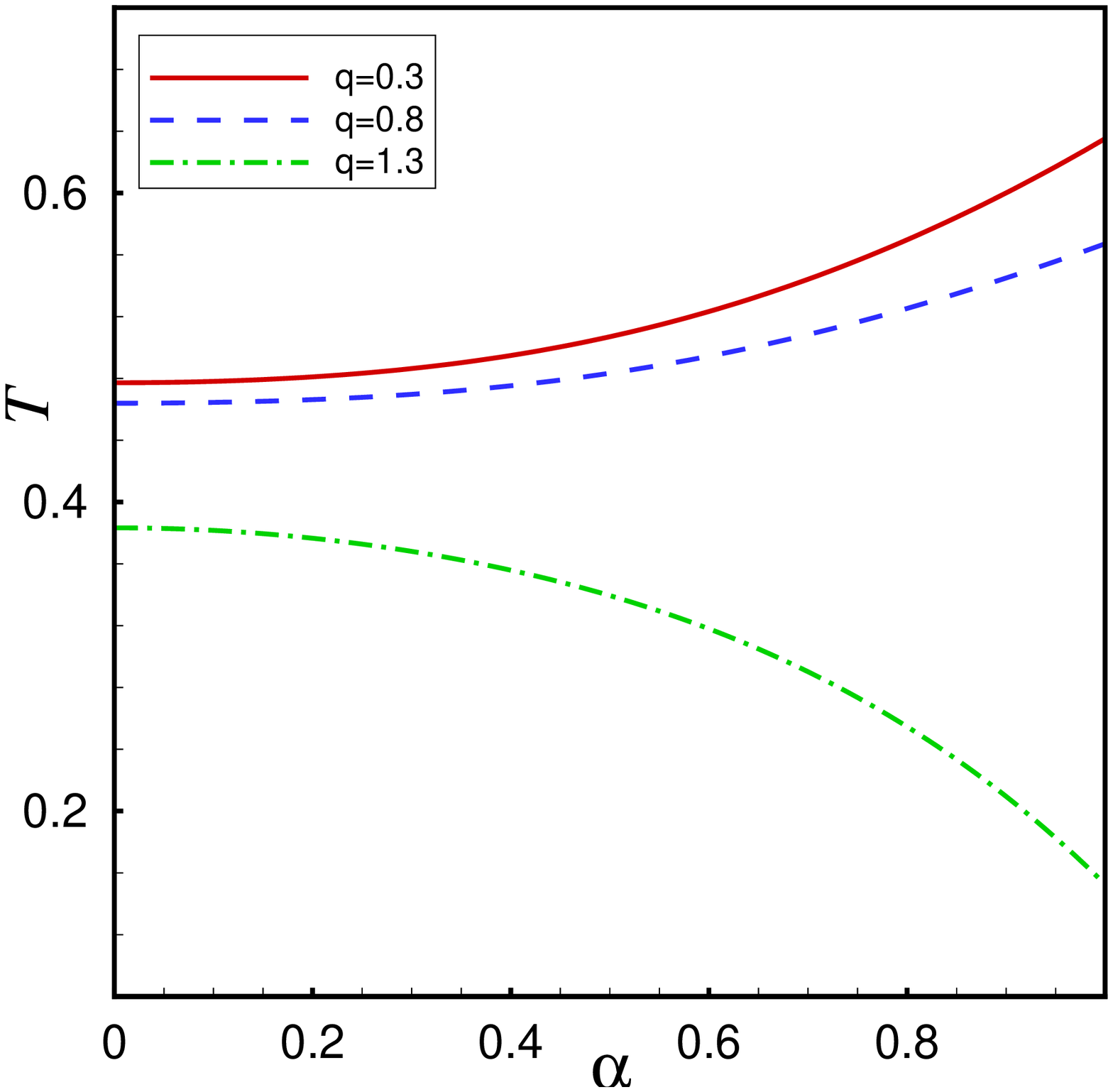}}
\caption{The behavior of $T$ versus $\protect\alpha \leq 1$ with $l=b=1$, $%
r_{+}=1.5$, $\Xi =1.25$, $n=5$ and $p=2$. }
\label{fig7}
\end{figure}

\begin{figure}[h]
\epsfxsize=7cm \centerline{\epsffile{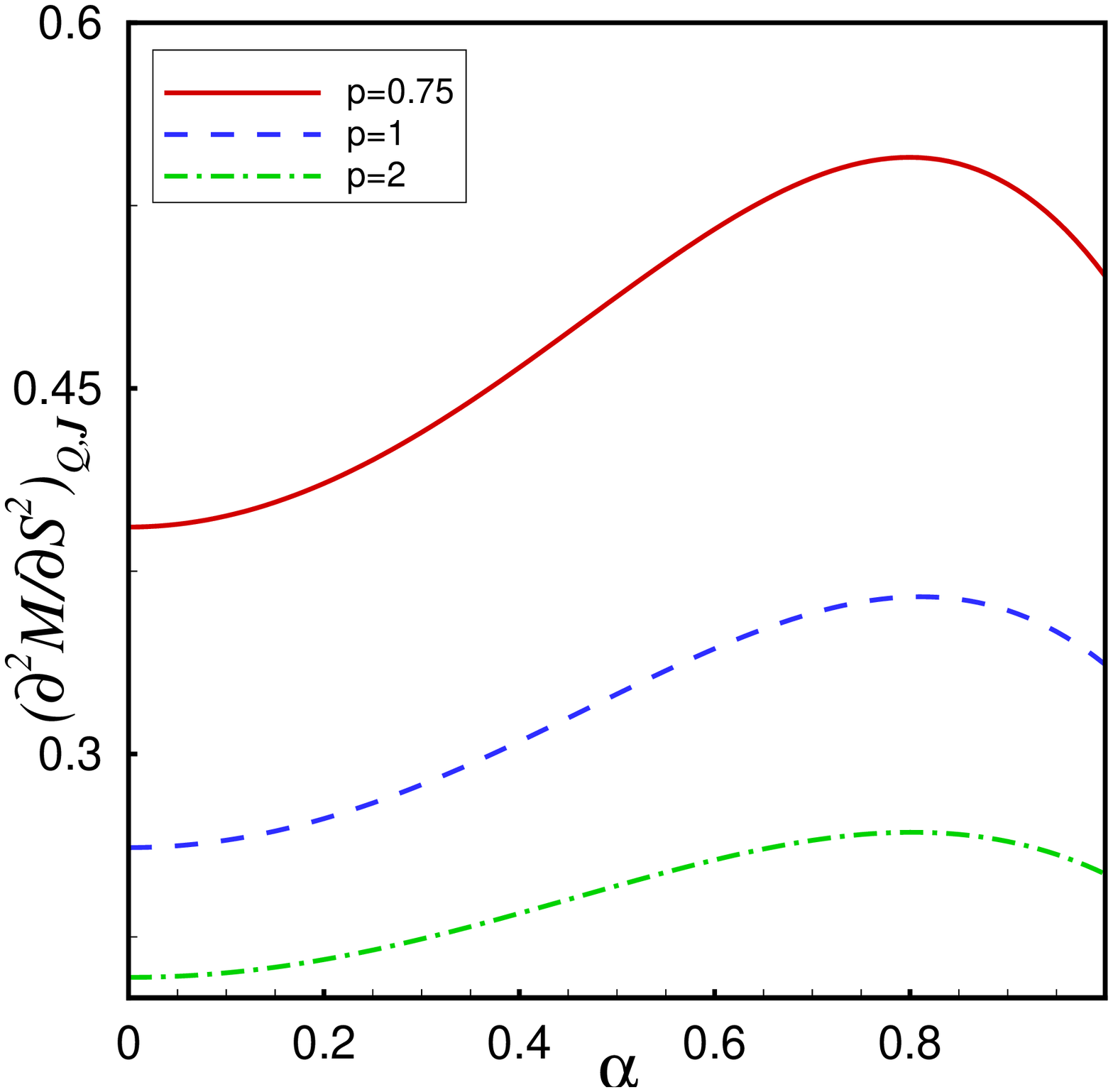}}
\caption{The behavior of $(\partial ^{2}M/\partial S^{2})_{Q,\mathbf{J}}$
versus $\protect\alpha \leq 1$ with $l=b=1$, $r_{+}=1.5$, $\Xi =1.25$, $n=5$
and $q=0.8$. Note that curve corresponding to $p=0.75$ has been rescaled by
the factor $1.5$.}
\label{fig8}
\end{figure}

\begin{figure}[h]
\epsfxsize=7cm \centerline{\epsffile{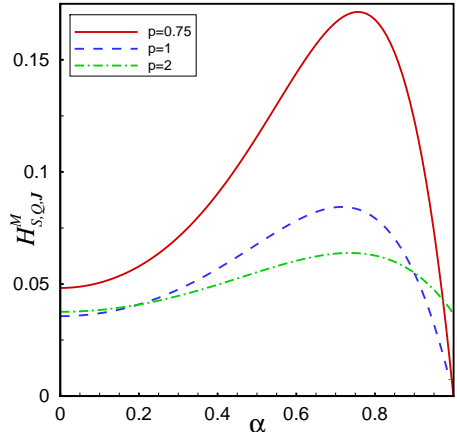}}
\caption{The behavior of $\mathbf{H}_{SQ\mathbf{J}}^{M}$ versus $\protect%
\alpha \leq 1$ with $l=b=1$, $r_{+}=1.5$, $\Xi =1.25$, $n=5$ and $q=0.8$.
Note that curves corresponding to $p=0.75$ and $p=2$ have been rescaled by
the factors $10$ and $10^{-1}$ respectively.}
\label{fig9}
\end{figure}

\begin{figure}[h]
\epsfxsize=7cm \centerline{\epsffile{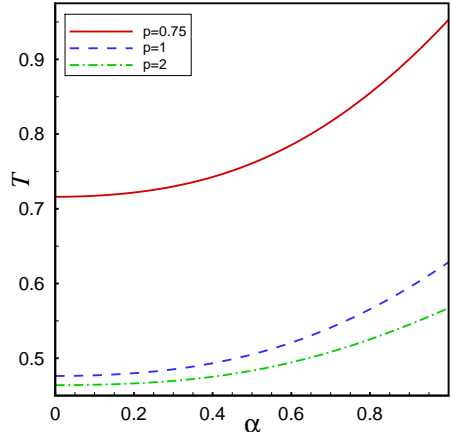}}
\caption{The behavior of $T$ versus $\protect\alpha \leq 1$ with $l=b=1$, $%
r_{+}=1.5$, $\Xi =1.25$, $n=5$ and $q=0.8$. Note that curve corresponding to 
$p=0.75$ has been rescaled by the factor $1.5$.}
\label{fig10}
\end{figure}

\begin{figure}[h]
\epsfxsize=7cm \centerline{\epsffile{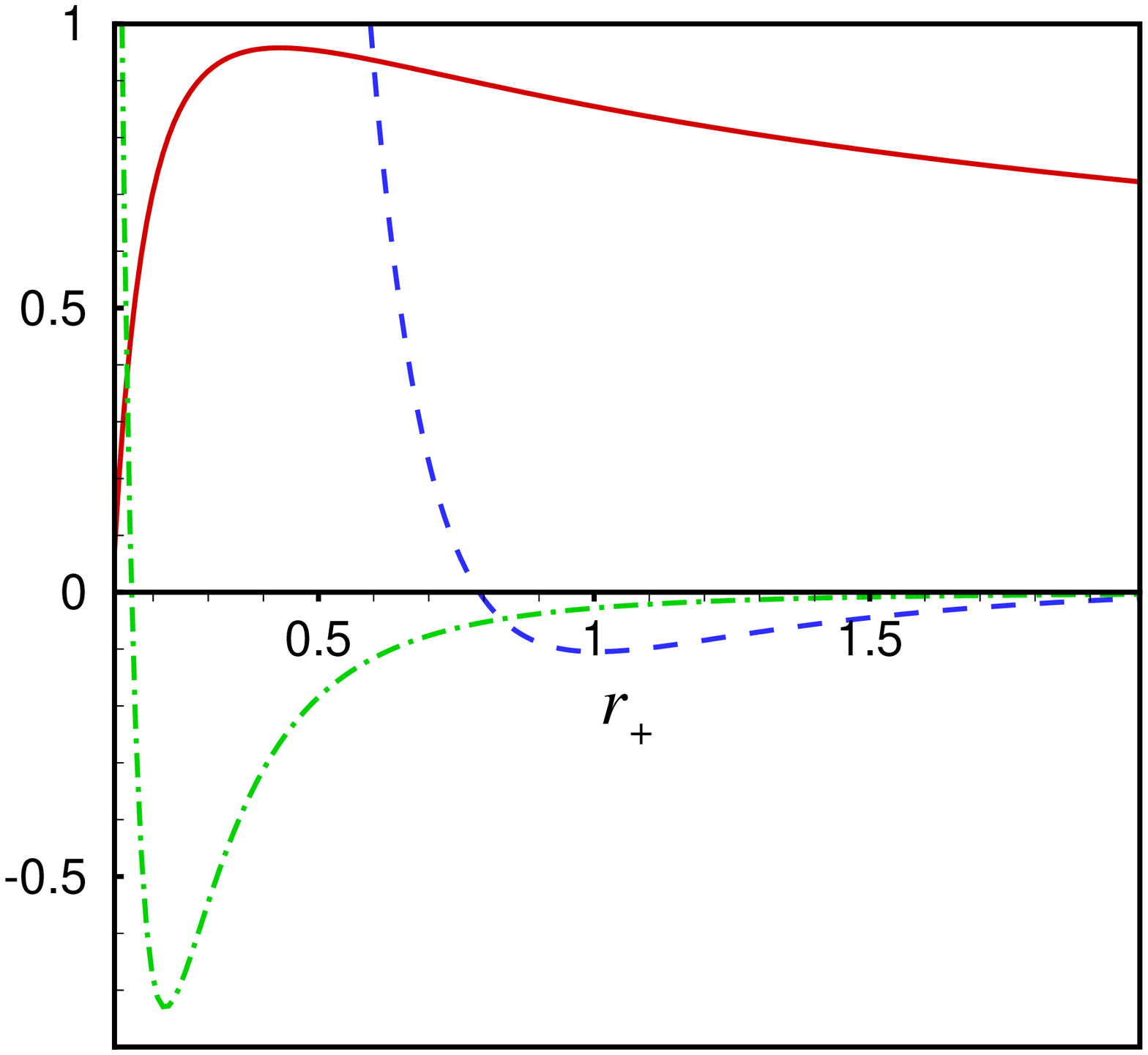}}
\caption{The behavior of $T$ (solid curve), $10(\partial ^{2}M/\partial
S^{2})_{Q,\mathbf{J}}$ (dashed curve) and $10^{-3}\mathbf{H}_{SQ\mathbf{J}%
}^{M}$ (dashdot curve) versus $r_{+}$ with $l=b=1$, $q=0.5$, $\protect\alpha %
=1.35$, $\Xi =1.25$, $n=5$ and $p=2$.}
\label{fig11}
\end{figure}

\begin{figure}[h]
\epsfxsize=7cm \centerline{\epsffile{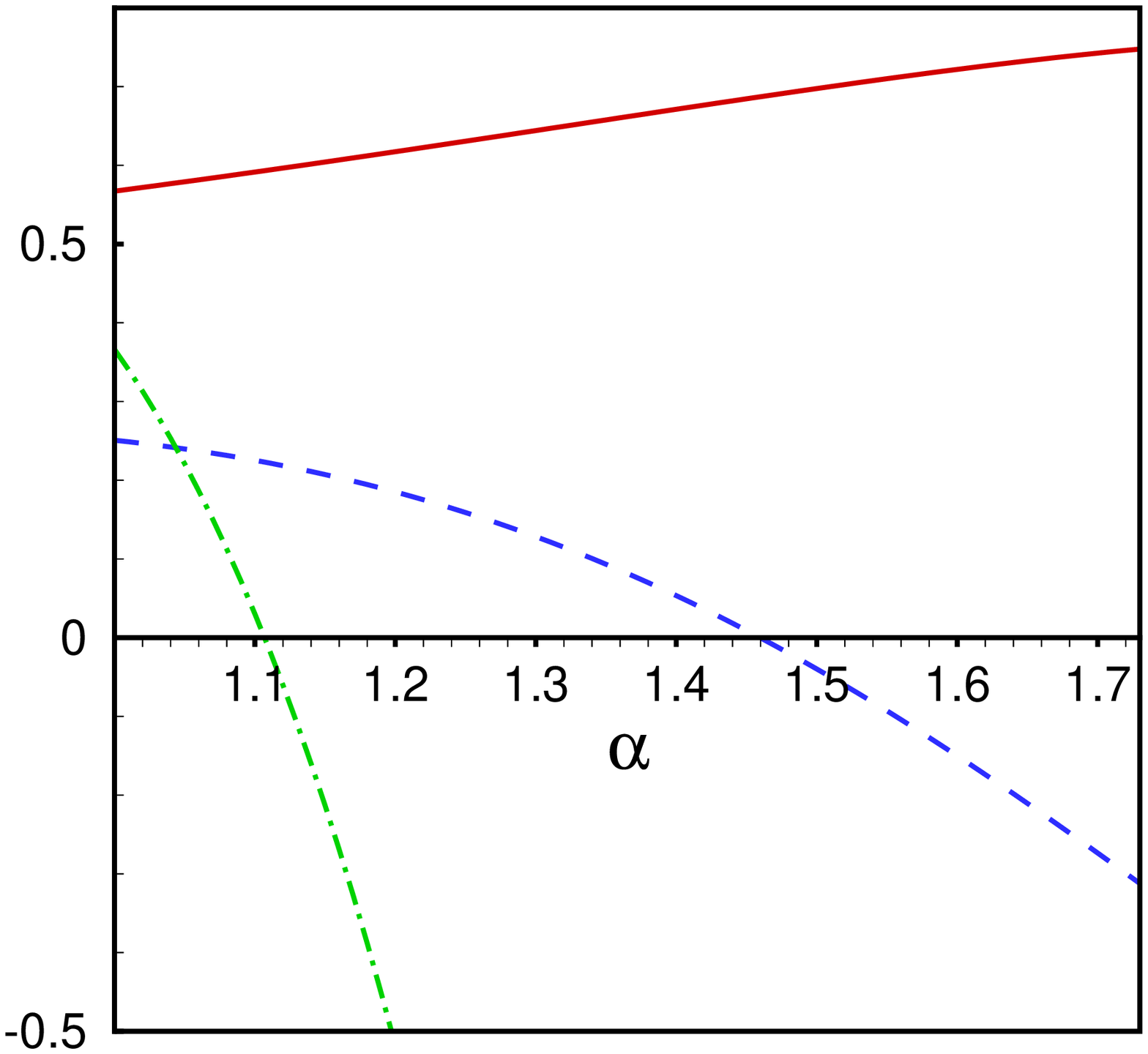}}
\caption{The behavior of $T$ (solid curve), $(\partial ^{2}M/\partial
S^{2})_{Q,\mathbf{J}}$ (dashed curve) and $\mathbf{H}_{SQ\mathbf{J}}^{M}$
(dashdot curve) versus $\protect\alpha $ with $l=b=1$, $q=0.8$, $r_{+}=1.5$, 
$\Xi =1.25$, $n=5$ and $p=2$.}
\label{fig12}
\end{figure}

\begin{figure}[h]
\epsfxsize=7cm \centerline{\epsffile{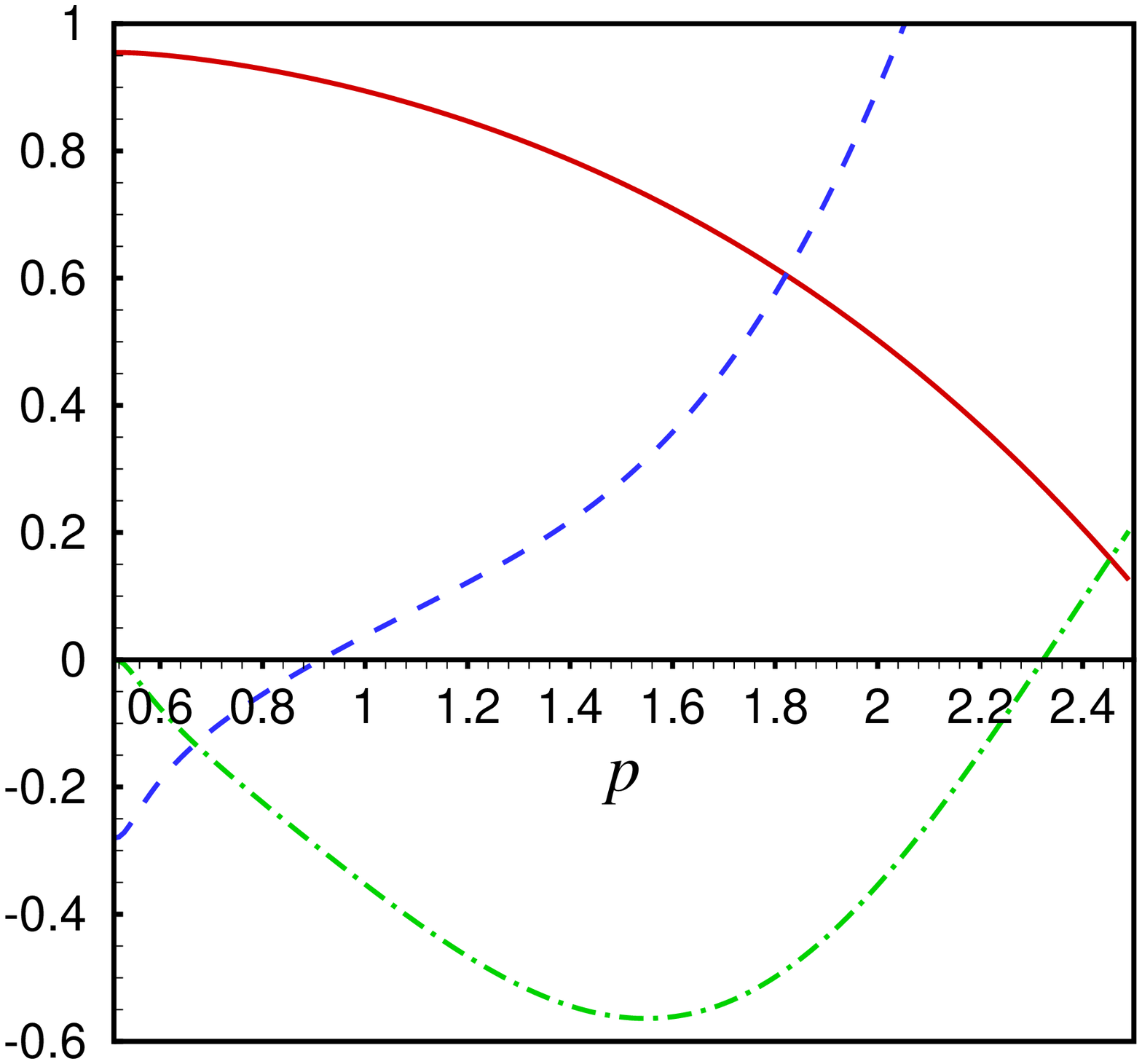}}
\caption{The behavior of $T$ (solid curve), $(\partial ^{2}M/\partial
S^{2})_{Q,\mathbf{J}}$ (dashed curve) and $10^{-1}\mathbf{H}_{SQ\mathbf{J}%
}^{M}$ (dashdot curve) versus $p$ with $l=b=1$, $q=0.9$, $r_{+}=1.1$, $\Xi
=1.25$, $\protect\alpha =1.45$ and $n=5$.}
\label{fig13}
\end{figure}

\begin{figure}[h]
\epsfxsize=7cm \centerline{\epsffile{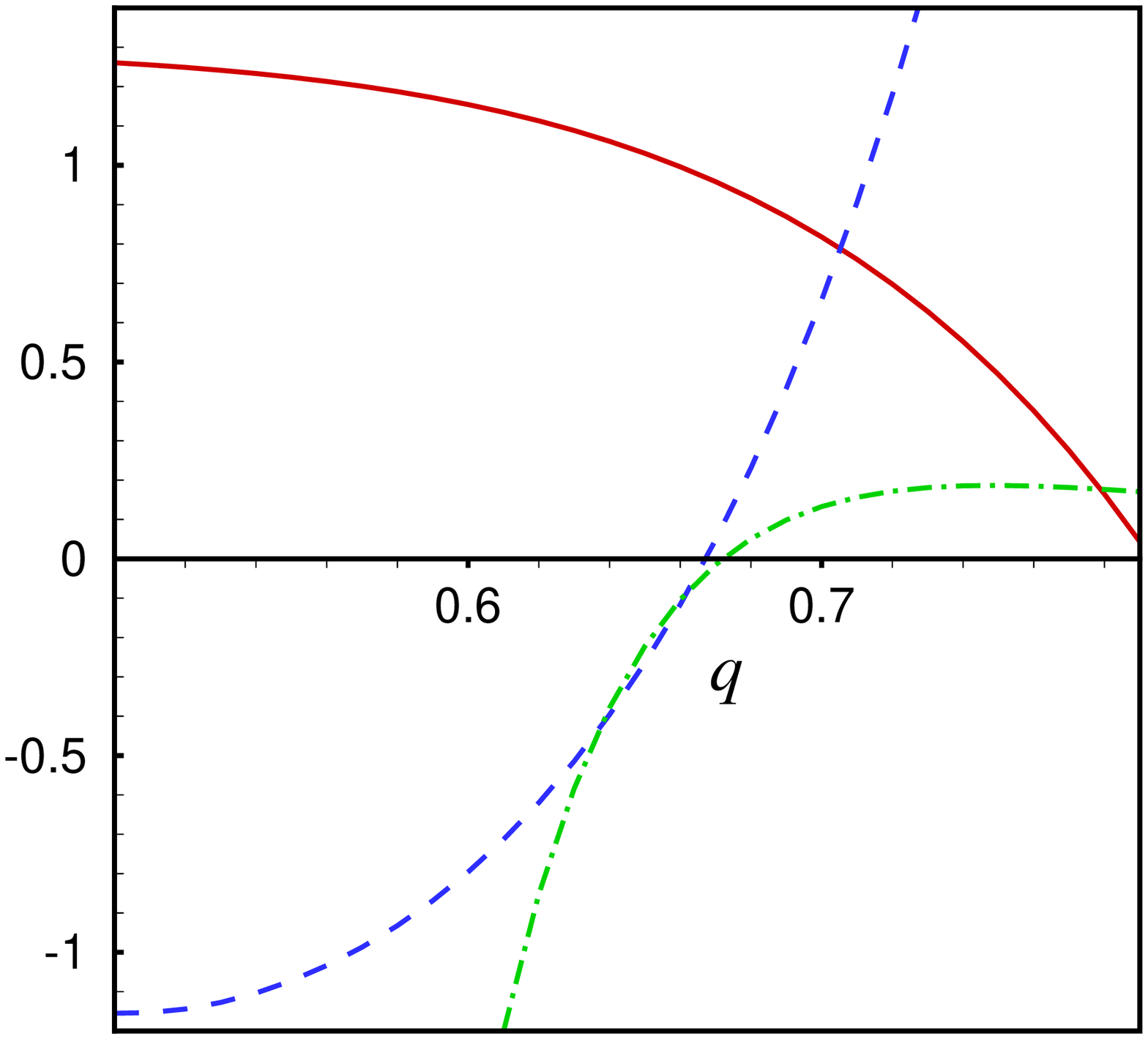}}
\caption{The behavior of $T$ (solid curve), $(\partial ^{2}M/\partial
S^{2})_{Q,\mathbf{J}}$ (dashed curve) and $10^{-2}\mathbf{H}_{SQ\mathbf{J}%
}^{M}$ (dashdot curve) versus $q$ with $l=b=1$, $p=4$, $r_{+}=0.9$, $\Xi
=1.25$, $\protect\alpha =1.5$ and $n=6$.}
\label{fig14}
\end{figure}

In this section, we are going to study thermal stability of rotating black
branes in the presence of a power-law Maxwell field in canonical and
grand-canonical ensembles. It is necessary for a thermodynamic system to
discuss its thermal stability. Thermal stability is investigated to ensure
that the entropy of system is at local maximum or equivalently the internal
energy of system is at local minimum. Therefore, the stability of a rotating
black brane as a thermodynamic system can be studied in terms of entropy $%
S(M,Q,\mathbf{J})$\ or its Legendre transformation $M(S,Q,\mathbf{J})$. In
terms of mass $M(S,Q,\mathbf{J})$, the local stability in any ensemble
implies that $M(S,Q,\mathbf{J})$ be a convex function of its extensive
variables. Typically, this behavior is studied by calculating the
determinant of the Hessian matrix of $M(S,Q,\mathbf{J})$\ with respect to
its extensive variables $X_{i}$, $\mathbf{H}_{X_{i}X_{j}}^{M}=\left[
\partial ^{2}M/\partial X_{i}\partial X_{j}\right] $\ \cite{Cal,Gub}. $%
\mathbf{H}_{X_{i}X_{j}}^{M}\geq 0$ guarantees that the system is thermally
stable. The number of thermodynamic variables is ensemble-dependent. In
canonical ensemble, the charge and angular momenta are fixed parameters and
consequently the determinant of Hessian matrix $\mathbf{H}_{X_{i}X_{j}}^{M}$
reduces to $(\partial ^{2}M/\partial S^{2})_{Q,\mathbf{J}}$. Therefore, in
order to find the ranges where the system is at thermal stability, it is
sufficient to find the ranges of positivity of $(\partial ^{2}M/\partial
S^{2})_{Q,\mathbf{J}}$ where the temperature $T$ is positive as well. In
grand-canonical ensemble $Q$\ and $\mathbf{J}$\ are not fixed parameters.

We discuss thermal stability for uncharged and charged cases, separately.
First we discuss the uncharged case. It is notable to mention that in the
case of uncharged rotating black branes $(\partial ^{2}M/\partial S^{2})_{%
\mathbf{J}}$ is exactly one which is obtained in \cite{Shey3}: 
\begin{equation}
\left( \frac{\partial ^{2}M}{\partial S^{2}}\right) _{\mathbf{J}}=\frac{%
n\left( {\alpha }^{2}+1\right) \left[ ({\Xi }^{2}-1)(n+1-2{\alpha }^{2})+{%
\Xi }^{2}\left( 1-{\alpha }^{2}\right) \right] }{{\pi \Xi }^{2}{l}^{4-n}{b}^{%
{\left( n-3\right) \gamma }}\left[ ({\alpha }^{2}+n-2){\Xi }^{2}+1-{\alpha }%
^{2}\right] }{r}_{+}^{{(2-n-{\alpha }^{2})/(\alpha }^{2}+1)}.  \label{ddMS}
\end{equation}%
Therefore, in canonical ensemble we just briefly review the results of them.
Since $\Xi ^{2}\geq 1$, (\ref{ddMS}) is positive provided $\alpha \leq 1$.
Therefore uncharged rotating black branes are stable in canonical ensemble
for $\alpha \leq 1$. Note that for uncharged case temperature is always
positive (see Eq. (\ref{Tem})). In grand-canonical ensemble $\mathbf{H}_{S%
\mathbf{J}}^{M}$ can be calculated as: 
\begin{equation}
\mathbf{H}_{S\mathbf{J}}^{M}=\frac{16\left( 1-\,{\alpha }^{2}\right) {l}%
^{2(n-3)}{r}_{+}^{2{(1-\,n)/(\alpha }^{2}+1)}}{{b}^{{2\left( n-1\right)
\gamma }}{\Xi }^{4}\left[ \left( {\alpha }^{2}+n-2\right) {\Xi }^{2}+1-{%
\alpha }^{2}\right] }.  \label{HMQJ}
\end{equation}%
From Eq. (\ref{HMQJ}) it is obvious that $\mathbf{H}_{S\mathbf{J}}^{M}\geq 0$
provided $\alpha \leq 1$, which is similar to the case of the canonical
ensemble. From the above arguments we conclude that the uncharged rotating
black branes are thermally stable provided $\alpha \leq 1$ in both canonical
and grand-canonical ensembles.

For charged case, we discuss the stability for $\alpha \leq 1$ and $\alpha
>1 $, separately. Since charge does not change the stable solutions to
unstable ones \cite{Deh3}, for $\alpha \leq 1$\ we have thermally stable
charged rotating black branes. This fact is shown in Figs. \ref{fig2}, \ref%
{fig3}, \ref{fig5}, \ref{fig6}, \ref{fig8} and \ref{fig9}. Figs. \ref{fig2}
and \ref{fig3} show that for $\alpha \leq 1$, the obtained solutions are
always stable in canonical and grand-canonical ensembles for any value of $%
r_{+}$, as we expect. Since $T>0$ guarantees that we have black branes for
our choices, one should choose $q<q_{\mathrm{ext}}$. The behavior of
temperature is depicted in Fig. \ref{fig4} with same constants as Figs. \ref%
{fig2} and \ref{fig3}. Behaviors of $(\partial ^{2}M/\partial S^{2})_{Q,%
\mathbf{J}}$ and $\mathbf{H}_{SQ\mathbf{J}}^{M}$ with respect to $\alpha
\leq 1$ for different choices of charge $q$ are plotted in Figs. \ref{fig5}
and \ref{fig6}. These figures once again show that charge cannot change
stable solutions to unstable ones and therefore we have stable charged
rotating solutions for $\alpha \leq 1$ as we had in uncharged case. Fig. \ref%
{fig7} shows positivity of temperature $T$ for solutions that their thermal
stabilities have been depicted in Figs. \ref{fig5} and \ref{fig6}. Figs. \ref%
{fig8} and \ref{fig9} depict stability of solutions for $\alpha \leq 1$ for
different values of $p$ in canonical and grand-canonical ensembles
respectively. The behavior of temperature for latter case is illustrated in
Fig. \ref{fig10}. For $\alpha >1$, one can understand from Fig. \ref{fig11}
that event horizon radius of stable black branes encounter an upper limit $%
r_{+\max }$ in both canonical and grand-canonical ensembles. The value of
this upper limit is greater in canonical ensemble than grand-canonical
ensemble. The effect of $\alpha $ on stability of solutions in both
canonical and grand-canonical ensembles are depicted in Fig. \ref{fig12}.
There is again an ensemble dependent upper limit this time on $\alpha $ that
for values greater than it solutions are no longer stable. The value of $%
\alpha _{\max }$ is smaller in the grand-canonical ensemble. In terms of $%
1/2<p<n/2$, stability is shown in Fig. \ref{fig13}. In contrast with $r_{+}$
and $\alpha $, there is a lower limit for $p$ i.e. for $p>p_{\min }$, we
have stable solutions. $p_{\min }$ is again ensemble dependent as well as $%
r_{+\max }$ and $\alpha _{\max }$. For $n/2<p<n-1$\ where $\alpha $\ has a $%
p $-dependent lower limit, numerical analysis confirm the result of
investigation for $1/2<p<n/2$, i.e. there is again a lower limit $p_{\min }$
that for values lower than it solutions are unstable. Stability in terms of
charge is depicted in Fig. \ref{fig14}. In this case there is an ensemble
dependent $q_{\min }$\ in each of ensembles that for $q$\ greater than it
black branes are stable. The value of $q_{\min }$ is greater in
grand-canonical ensemble.

\section{CLOSING REMARKS}

In this paper, we constructed a new class of higher dimensional nonlinear
charged rotating black brane solutions in Einstein-dilaton gravity with
complete set of rotation parameters. The nonlinear electromagnetic source
was considered in the form of the power-law Maxwell field which guarantees
conformal invariance of the electromagnetic Lagrangian in arbitrary
dimensions for specific choices of power. Due to the presence of the dilaton
field, our solutions are neither asymptotically flat nor (A)dS. We showed
that in case of power-law Maxwell field, one needs two Liouville type
potentials in order to have a rotating black brane solutions while in the
case of linear Maxwell source just one term is needed \cite{Shey3}. The
extra dilaton potential term disappears for $p=1$. All our results reproduce
the results of \cite{Shey3} in the case of linear charged rotating solutions
where $p=1$.

Demanding from one side that the value of total mass should be finite and
from the other side that the effect of mass term in metric function should
disappear at infinity, we found some restrictions on $p$ and $\alpha $. The
allowed ranges of these two parameters is as follows. For $1/2<p<n/2$, we
have $0\leq \alpha ^{2}<n-2$ while for $n/2<p<n-1$, we have $2p-n<\alpha
^{2}<n-2$. For these permitted ranges, our solutions are always well-defined.
Also, in these ranges of $p$ and $\alpha $, the charge term in metric
function $f(r)$ is always positive and dominant in the vicinity of $r=0$.
Therefore, Schwarzschild-like solutions are ruled out. However, solutions
with two inner and outer horizons, extreme solutions and naked singularities
are allowed.

In order to study thermodynamics of charged rotating black branes, we
calculated mass, charge, temperature, entropy, electric potential energy and
angular momentum. Using these quantities we obtained Smarr-type formula for
the mass $M(S,Q,\mathbf{J})$ and showed that the first law of thermodynamics
is satisfied. Next, we analysed thermal stability of the solutions in both
canonical and grand-canonical ensembles. These investigations showed that
for $\alpha \leq 1$ charged rotating black brane solutions are always stable
with any value of the charge parameter. For $\alpha >1$, there is a $%
r_{+\max }$ in each of ensembles that we have stable solutions provided
their radius are smaller than $r_{+\max }$. In terms of $\alpha (>1)$, the
solutions is changed from stable ones to unstable ones when they meet an $%
\alpha _{\max }$. Versus $p$ and $q$, we showed that the solutions encounter 
$p_{\min }$ and $q_{\min }$ respectively so that solutions with $q$ and $p$
parameters lower than them are unstable for $\alpha >1$. All $r_{+\max }$, $%
\alpha _{\max }$, $q_{\min }$ and $p_{\min }$ values depend on the ensemble.

It is worth noting that the higher dimensional charged rotating solutions
obtained here have flat horizon. One may interested in studying the rotating
solutions with curved horizon. Specially the case of spherical horizon will
be a good extension of Kerr-Newmann solution. It seems that the study of the
general case is a difficult problem. However, it is possible to seek for
slowly rotating nonlinear charged solutions with curved horizon. The latter
is in progress.

\acknowledgments{We thank Shiraz University Research Council. This
work has been supported financially by Research Institute for
Astronomy \& Astrophysics of Maragha (RIAAM), Iran.}

\end{document}